\documentclass[12pt]{article}
\usepackage[english]{babel}
\usepackage{graphicx} 
\usepackage{amsmath,amsthm}
\usepackage{amsfonts}
\usepackage[dvips]{rotating}

\begin{document}

\title{On an economic prediction of the finer resolution level wavelet coefficients in electron structure calculations}%
\author{Szilvia Nagy$^a$ and J\'anos Pipek$^b$}%
\thanks{\texttt{nagysz@sze.hu}$^a$Sz\'echenyi Istv\'an University, Gy\H{o}r, Hungary, nagysz@sze.hu, $^b$Budapest University of Technology and Economics, Budapest, Hungary, pipek@phy.bme.hu}%
\date{\today}%
\begin{abstract}
In wavelet based electron structure calculations introducing a new, finer resolution level is usually an expensive task, this is why often a two-level approximation is used with very fine starting resolution level. This process results in large matrices to calculate with and a large number of coefficients to be stored. In our previous work we have developed an adaptively refining solution scheme that determines the indices, where refined basis functions are to be included, and later a method for predicting the next, finer resolution coefficients in a very economic way. In the present contribution we would like to determine, whether the method can be applied for predicting not only the first, but also the other, higher resolution level coefficients. Also the energy expectation values of the predicted wave functions are studied, as well as the scaling behaviour of the coefficients in the fine resolution limit.

\end{abstract}
\maketitle
\section{Introduction}

In data analysis wavelets\cite{Daub} have a large share within the methods both in scientific and industrial research and applications. The theory of using wavelet basis functions offers two possible, different interpretations. The analytic aspect reduces the decomposition of a function, image or signal to more and more simple (more and more rough) building blocks as the analysis proceeds. On the other hand, the synthetic point of view starts from a rough resolution level of a function and by adding refinements, it arrives at a sufficiently precise representation of the function that is studied.

The most widespread use of wavelets is still in the image compression techniques\cite{Skodras, ICER}, and the other applications are also mainly analyzers. Wavelets can build a basis for differential equation discretisation\cite{Galerkin} and solving\cite{Urban, Dahmen, dahmen2}, and the solvers have been developed and tested in various fields of science from diffusions to electromagnetic waves\cite{Nasif,D1,D2, ATJ}. In electron structure calculations wavelet basis has been present since the early nineties\cite{Arias, Goed, GoedRMP, Harrison}, and in the previous decade both a wavelet based\cite{Goed, GoedRMP, Genovese} and a multiwavelet based\cite{Fredian1, Fredian2, Fredian3} solver have been developed with chemical accuracy and massively parallel computation possibility. These solvers mostly use two resolution levels, but adaptively refining solution schemes are also given for simpler systems\cite{Flad,ot,TCA}.

As the wavelet basis set behaves as a set of building blocks that can be chosen uniformly, independently of the system itself, we have studied the coefficients of the electron-electron cusp in the two-electron density matrix\cite{ket}, so that the electron-electron cusp could be added to a rougher resolution level solution as a last refinement step. In the following a similar idea is presented, a prediction of the next level coefficients is suggested\cite{jcc}, moreover, the asymptotic behavior of the components of the predicted wavelet coefficients, the predicted energy levels, and the possibility of using the computationally cheap prediction for further refinements are studied based on the suggestions from Ref.\cite{Deutsch}.

\subsection{About wavelet analysis}
In order to introduce the notations we use in the article for the wavelets, we shortly summarize the idea behind wavelet analysis. In the discrete wavelet analysis the Hilbert space of the problem to be solved is divided into resolution levels which are embedded into each other. The basis functions of each resolution level consist of shifted versions of one function on a regular grid, and the grid distance halves at each consecutive resolution levels. Let us use the notation for the $m$th resolution level scaling function
\begin{equation}\label{scfun}
    s_{m,k}(x)=2^{m/2} s(2^m x-k)
\end{equation}
where $s(x)$ is the ``mother scaling function'', and $k$ is the shift index. As the refinement levels are embedded into one another, i.e., any function that can be exactly expanded at resolution level $m$ can be also exactly expanded at any finer resolution level $m+n$. In particular, the mother scaling function belonging to the resolution level $m=0$ can be expanded by the scaling functions of level $m=1$, i.e. there exists a refinement equation between the neighboring resolution level scaling functions,
\begin{equation}\label{refineeqs}
    s(x)=2^{1/2}\sum_{i=0}^{N_s}h_i s(2x-i),
\end{equation}
with $\sum_{i=0}^{N_s}h_i=\sqrt{2}$. Wavelets are the basis functions of the detail space which completes a rough resolution subspace to the next, refined subspace of the Hilbert space. They are also shifted and shrunk versions of one common ``mother wavelet'' $w(x)$,
\begin{equation}\label{wfun}
    w_{m,k}(x)=2^{m/2} w(2^m x-k).
\end{equation}
The mother wavelet, as it is an element of the subspace $m=1$ can also be expanded by scaling functions,
\begin{equation}\label{refineeqw}
    w(x)=2^{1/2}\sum_{i=0}^{N_s}g_i s(2x-i),
\end{equation}
with $g_i=(-1)^ih_{N_s-i}$.

In order to simplify later usage of the basis functions, a general
\begin{equation}\label{basis_chi}
    \chi_\tau(x)=\left\lbrace \begin{array}{c}
                             s_{m,k}(x),\quad \mbox{if}\ \tau=\lbrace s,m,k\rbrace \\
                             w_{m,k}(x),\quad \mbox{if}\ \tau=\lbrace w,m,k\rbrace
                           \end{array}\right.
\end{equation}
basis function with a composite index $\lambda$ will be introduced.

\section{Prediction of the first finer resolution level coefficients in the wavelet-based solution of the Schr\"odinger equation}

A wave function can be expanded at a given resolution level $M$ either as linear combination of the scaling functions of the resolution level
 \begin{equation}\label{PSIdecomps}
  \Psi^{[M]}(x)=\sum_{\ell\in\Omega_M} c_{M\ell}\; s_{M\ell}(x),
 \end{equation}
or starting from a basic resolution level $m=0$
\begin{equation}\label{PSIdecompw}
    \Psi^{[M]}(x)=\sum_{\ell\in\Omega_0} c_{0\ell}\; s_{0\ell}(x)+\sum_{m=0}^{M-1} \sum_{\ell\in\Omega_m} d_{m\ell}\; w_{m\ell}(x).
\end{equation}
In both of the above cases $\Omega_m$ denotes the set of the non-zero expansion coefficients at resolution level $m$.

Using expansion~(\ref{PSIdecompw}), the Schr\"odinger equation
\begin{equation}\label{Sch}
   \hat{H}\Psi=E\Psi
\end{equation}
can be approximated at any resolution level $M$, resulting in
\begin{equation}\label{Sch_M}
   H^{[M]}\Psi^{[M]}=E^{[M]}\Psi^{[M]},
\end{equation}
with the matrix elements
\begin{equation}\label{Hamiltonmx_M}
   H^{[M]}_{\rho\tau}=\langle \chi_\rho|\hat{H}|\chi_\tau\rangle.
\end{equation}
Here the notations of~(\ref{basis_chi}) were used with $\chi$'s being either 0th resolution level scaling functions or wavelets of any resolution level  $m=0,1,\ldots,M-1$.

As we have mentioned previously, the point to use wavelets in solving differential equations is that most of the higher resolution level wavelet coefficients are close to zero, thus they can be omitted from the calculations. In Ref.~\cite{jcc} we suggested a method for predicting which of the next refinement level coefficients will be necessary to include to the refined calculation if the precision of the $M$th resolution level is not sufficient, and telling whether the $M$th resolution level sufficiently precise is by predicting the magnitude of the next resolution level wavelet coefficients. The method is summarized shortly in the followings.

Let us suppose, that we have solved the $M$th resolution level problem (\ref{Sch_M}) and having determined both $\Psi^{[M]}$, and the domains of the non-zero coefficients $\Omega_m$ at each resolution level $0\leq m<M$. As an approximation for the magnitude of the next level coefficients $d_{M,k}$, we can optimize the energy by adding just one wavelet from the subspace $M$ to the wave function. The new wave function
\begin{equation}\label{Phi_1wavelet}
    \Phi^{[M+1]}(\alpha_k)=\Psi^{[M]}+\alpha_k\cdot w_{M,k}
\end{equation}
results in a new energy
\begin{equation}\label{Energy1}
    \mathcal{E}(\alpha_k)=\frac{\langle \Phi^{[M+1]}(\alpha_k)|\hat{H}|\Phi^{[M+1]}(\alpha_k)\rangle}{\langle \Phi^{[M+1]}(\alpha_k)|\Phi^{[M+1]}(\alpha_k)\rangle}.
\end{equation}
Using the Ritz variation principle for the ground state results in
\begin{equation}\label{RitzE}
    \frac{d\mathcal{E}(\alpha_k)}{d\alpha_k}=0,\qquad \frac{d^2\mathcal{E}(\alpha_k)}{d\alpha_k^2}>0.
\end{equation}
The solution is
\begin{equation}\label{alfa}
  \alpha_k=\left\{\begin{array}{cc}
    -\lambda+\sqrt{\lambda^2+1}, &\quad\mbox{if } \langle w_{M k}|\hat{H}|\Psi^{[M]}\rangle> 0\\
    -\lambda-\sqrt{\lambda^2+1}, &\quad\mbox{if } \langle w_{M k}|\hat{H}|\Psi^{[M]}\rangle< 0\\
    0                     &\quad\mbox{if } \langle w_{M k}|\hat{H}|\Psi^{[M]}\rangle=0
  \end{array}\right.,
\end{equation}
where the shorthand notation
\begin{equation}\label{lambda}
  \lambda=\frac{E^{[M]}-\langle w_{M k}|\hat{H}|w_{M k}\rangle}{2\langle w_{M k}|\hat{H}|\Psi^{[M]}\rangle}
\end{equation}
was introduced. This value of $\alpha_k$ predicts the real value of the wavelet coefficients $d_{M,k}$ not only for the ground states, but also for the excited states as it was proven in Ref.~\cite{jcc}. It can also be seen, that the values of alpha are approximately $\alpha_k\approx\frac{1}{2\lambda}$ if $\langle w_{M k}|\hat{H}|\Psi^{[M]}\rangle$ is near zero.

\section{Matrix elements and singularities}

In the infinitely fine resolution limit the values $E^{[M]}=\langle \Psi^{[M]}|\hat{H}|\Psi^{[M]}\rangle$, $R_{M,k}=\langle w_{M k}|\hat{H}|\Psi^{[M]}\rangle$, and $W_{M,k}=\langle w_{M k}|\hat{H}|w_{M k}\rangle$ constituting the approximation $\alpha_k$ and their scaling properties with the resolution level $M$ can be calculated.

\begin{figure}[hbt]
\centering
  \includegraphics[height=6.5cm]{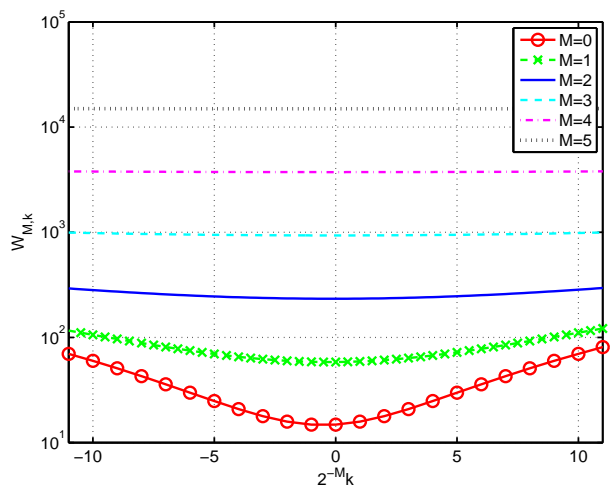}
  \caption{The values $W_{M,k}$ as a function of the normalized shift index $k$ for resolution levels $M=1,\ldots,6$ for a 1D harmonic oscillator model system with $\omega=1$~a.u., calculated with Daubechies basis set of support length 8. Atomic units were used.}
  \label{fig:W}
\end{figure}

The scaling properties of $W_{M,k}$ can be studied easily: it consists of the kinetic energy and the potential energy terms in one-electron systems. The generalization for multiple electrons can also be carried out. The scaling behaviour of kinetic energy term
\begin{equation}\label{WkinM}
    \langle w_{M,k}|-\frac{1}{2}\Delta|w_{M,k}\rangle
\end{equation}
with the resolution level $M$ can be calculated by using the definition
\begin{equation}\label{wavefunM}
    w_{M,k}=2^{\frac{M}{2}}w(2^Mx-k)
\end{equation}
and applying the variable transformation $y=2^Mx-k$, results in
\begin{equation}\label{Wkin_scale}
    \frac{1}{2}2^{2M}\langle w_{0,0}|\Delta|w_{0,0}\rangle.
\end{equation}
In case of the potential energy term
\begin{equation}\label{WpotM}
    \langle w_{M,k}|V|w_{M,k}\rangle=\int w_{M,k}^*(x)V(x)w_{M,k}(x) dx,
\end{equation}
the substitution (\ref{wavefunM}) and changing of the integration variable to $y=2^Mx-k$ result in
\begin{equation}\label{Wpot_scale}
\int w_{M,k}^*(y)V(2^{-M}(y+k))w_{M,k}(y) dy \rightarrow V(0)\langle w_{0,0}|w_{0,0}\rangle,
\end{equation}
if $M$ is large. The values of $W_{M,k}$ can be seen in Figure~\ref{fig:W}.
We have applied a one dimensional harmonic oscillator model, as the potential energy term, i.e., the expectation value of a second order polynomial can be exactly calculated using the results in Ref.\cite{DahmMicc}. Similarly, the potential terms of an electron confined in a box can be calculated according to Ref.\cite{ATJ}. The dominance of the kinetic energy term is clearly visible on the plot.

\begin{figure*}[hbt]
\centering
  \includegraphics[height=4.5cm]{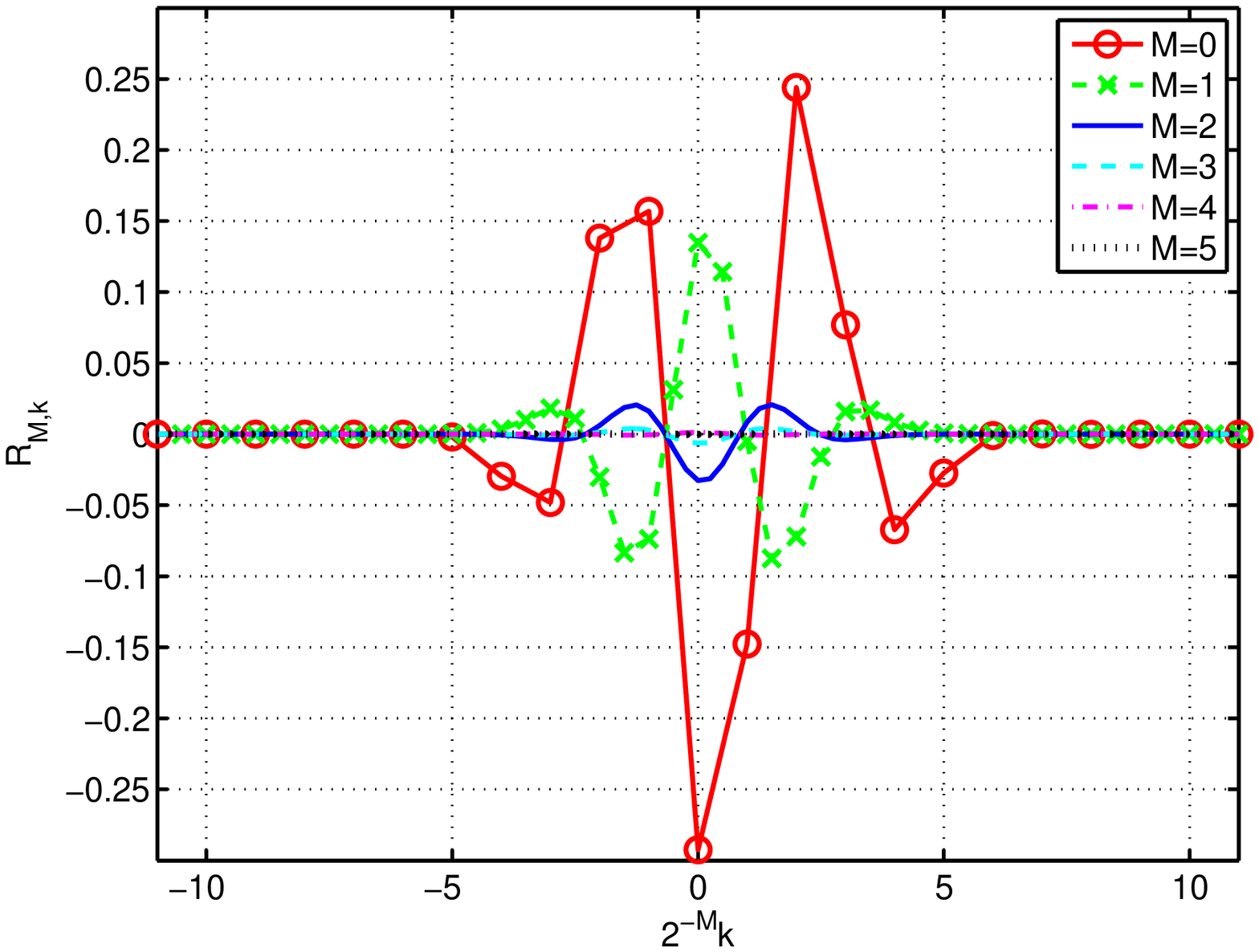}
  \includegraphics[height=4.65cm]{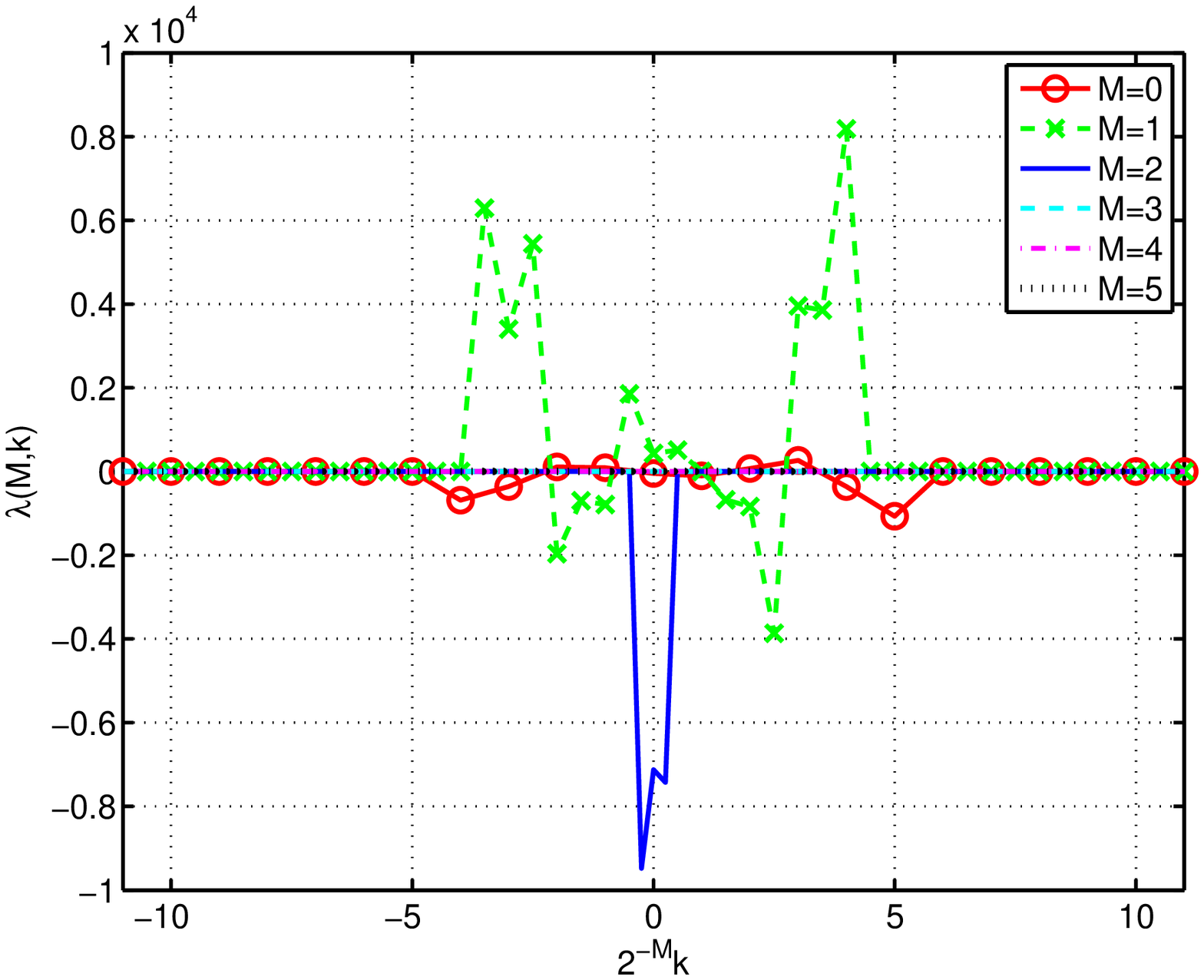}
  \includegraphics[height=4.5cm]{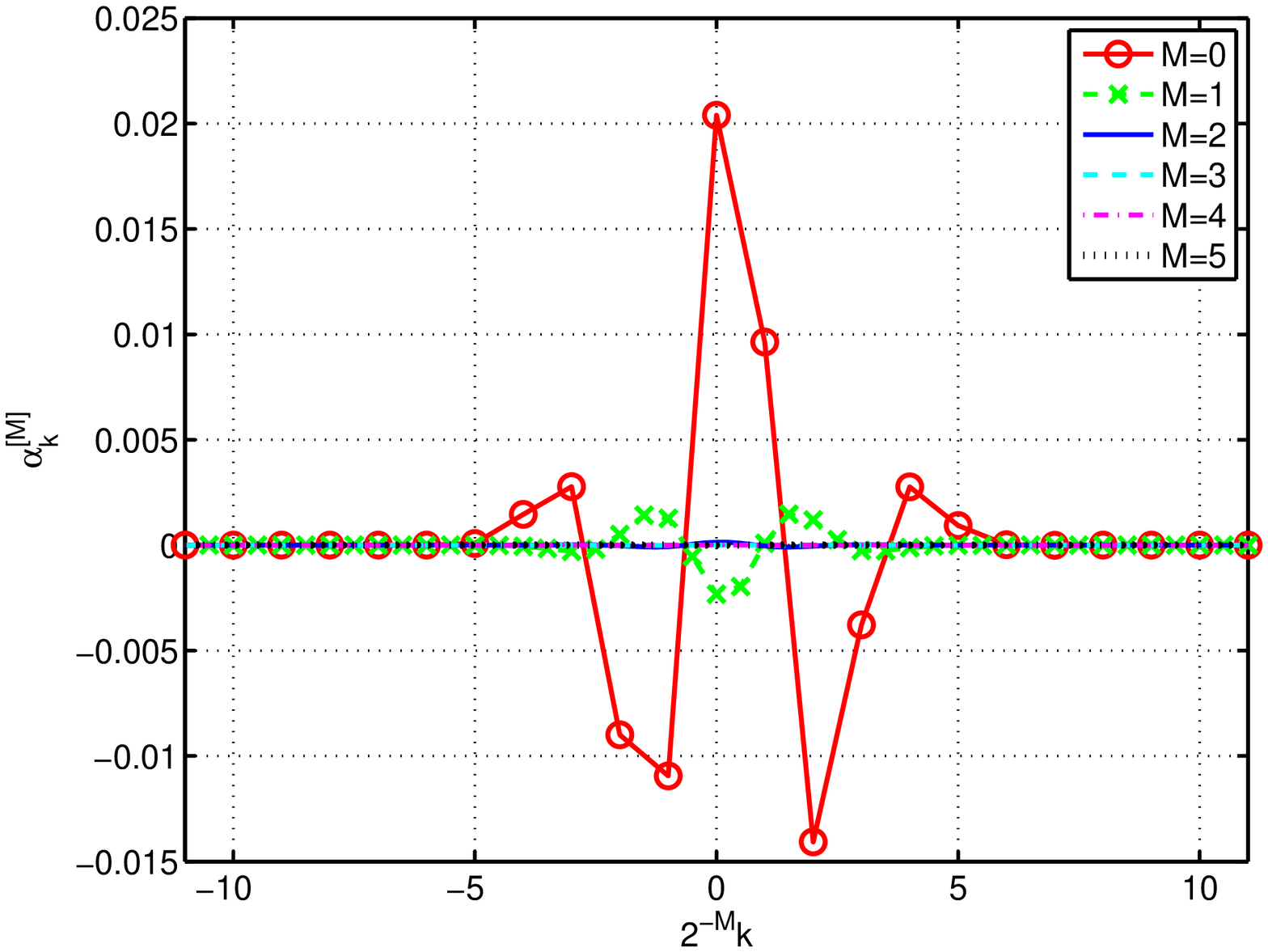}
  \caption{The values $R_{M,k}$, $\lambda$ and $\alpha_k$ as a function of the normalized shift index $k$ for resolution levels $M=1,\ldots,6$ for a 1D harmonic oscillator model system with $\omega=1$~a.u., calculated with Daubechies basis set of support length 8. Ground state and atomic units were used.}
  \label{fig:R}
\end{figure*}

Similar calculations can be carried out for determining the behavior of the values $R_{M,k}$. If $M$ is large enough, the value of $\hat{H}|\Psi^{[M]}\rangle$ in $R_{M,k}$ approximates $E|\Psi^{[\infty]}\rangle$ very well, as well as the energy $E^{[M]}$ approximates its infinitely fine resolution level limit, the exact energy $E$. Using these values, the limiting behavior of $R_{M,k}$ can be calculated as
\begin{equation}\label{RinfM}
    \langle w_{M,k}|E|\Psi^{[\infty]}\rangle=E\int w_{M,k}^*(x)\Psi^{[\infty]}(x) dx.
\end{equation}
A Taylor series expansion of the wave function around 0
\begin{equation}\label{PsiTaylor}
    \Psi(2^{-M}(y+k))\approx \Psi(0)+\Psi'(0)\cdot 2^{-M}(y+k)
\end{equation}
results in
\begin{equation}\label{Rinf_scale}
    2^{-\frac{3M}{2}}E\Psi'(0)\int y w^*(y)dy.
\end{equation}
Note, that the first moment of the wavelet, $\mu_1=\int y w^*(y)dy$ appears in the expression, as well as the exact energy and the derivative of the exact wave function at 0. The absolute values of $R_{M,k}$ can be seen in Figure~\ref{fig:R}.
The excited states as well as other basis sets result in very similar tendencies in $R_{M,k}$. The value of $R_{M,k}$ oscillates a lot, but the overall scaling, especially for larger $M$s meet the description in (\ref{Rinf_scale}). As it we have found previously\cite{ot,TCA}, the indices, where $R_{M,k}$ is large are the positions where further refinement might be necessary, moreover, for most of the indices $R_{M,k}$ behaves very similarly to $\alpha_k$.

The scaling properties of $\lambda$ can be easily derived from (\ref{Wpot_scale}) and (\ref{Rinf_scale}),
\begin{equation}\label{lambda_scale}
    \lambda \approx \frac{1}{2E\Psi'(0)}2^{\frac{7M}{2}}\frac{\langle w_{0,0}|\triangle|w_{0,0}\rangle}{\mu_1}.
\end{equation}
For large values of $\lambda$, i.e., when $R_{M,k}\approx0$, the predicted coefficients are approximately $\frac{1}{2\lambda}$, and the number of these $M$th level coefficients $\alpha_k$ is $2^M$ times the interval length, thus the total weight $2^M\alpha_k^2$ of the $M$th resolution level is less than $2^{-6M}$. If all the resolution levels above $M$ are omitted from the calculations, the total weight that is lost is less than
\begin{equation}\label{weight}
    \sum_{m=M}^\infty 2^{-6m}=\frac{2^{-6M}}{1-2^{-6}}.
\end{equation}

\section{Energy predictions}

\begin{figure}[hbt]
\centering
  \includegraphics[height=6.5cm]{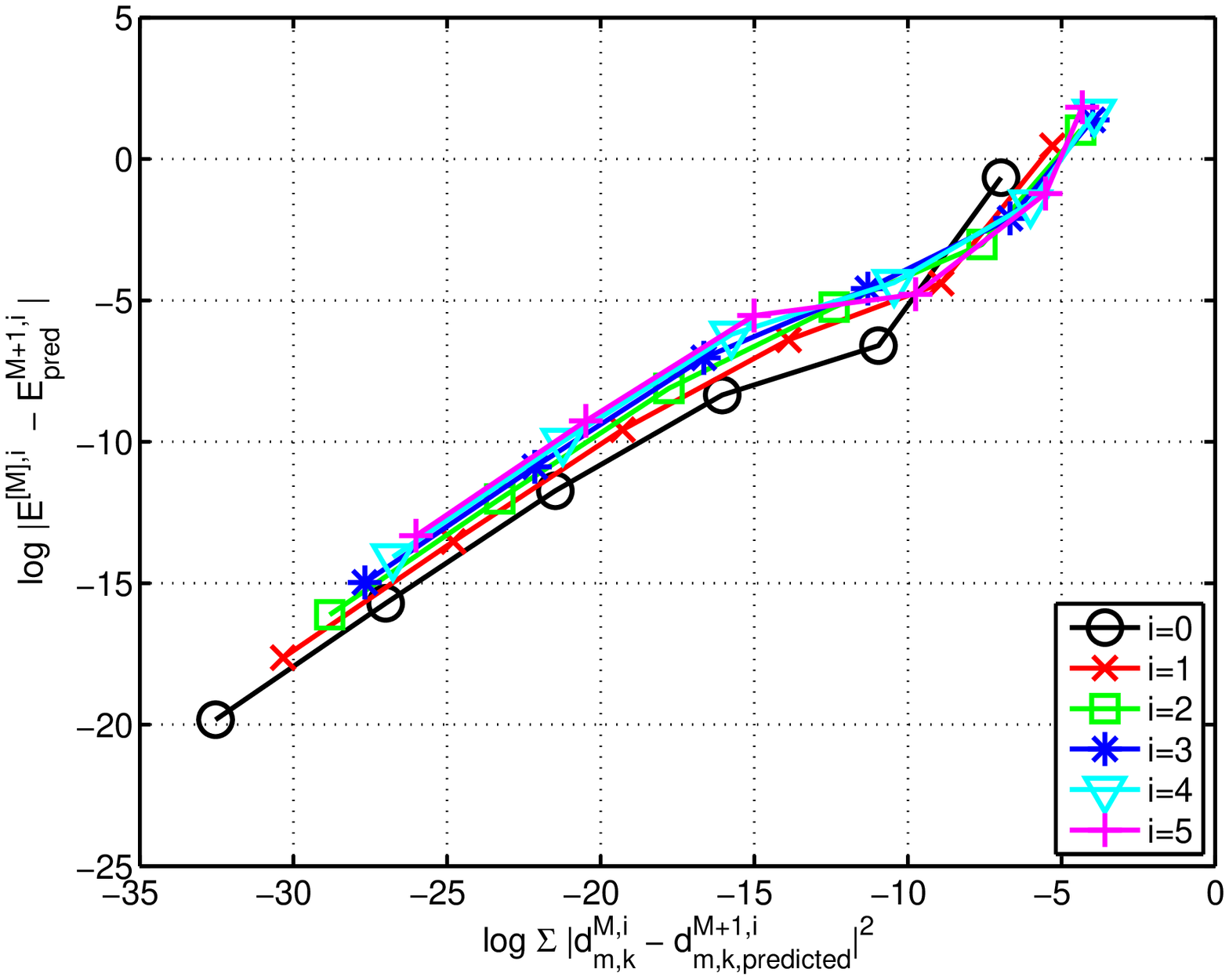}
  \caption{Energy errors vs. resolution level of a 1D harmonic oscillator model system with $\omega=1$~a.u., calculated with Daubechies basis set of support length 8. $E^{[M],i}$ denotes the energy levels calculated from the $M$th level eigenvalue equations and $E^{M,i}_{pred}$ means the $M$th level predictions calculated from the $M-1$th level solutions. The excitation index is $i$. Atomic units were used.}
  \label{fig:energdif}
\end{figure}

\begin{figure}[htb]
\centering
  \includegraphics[height=6.5cm]{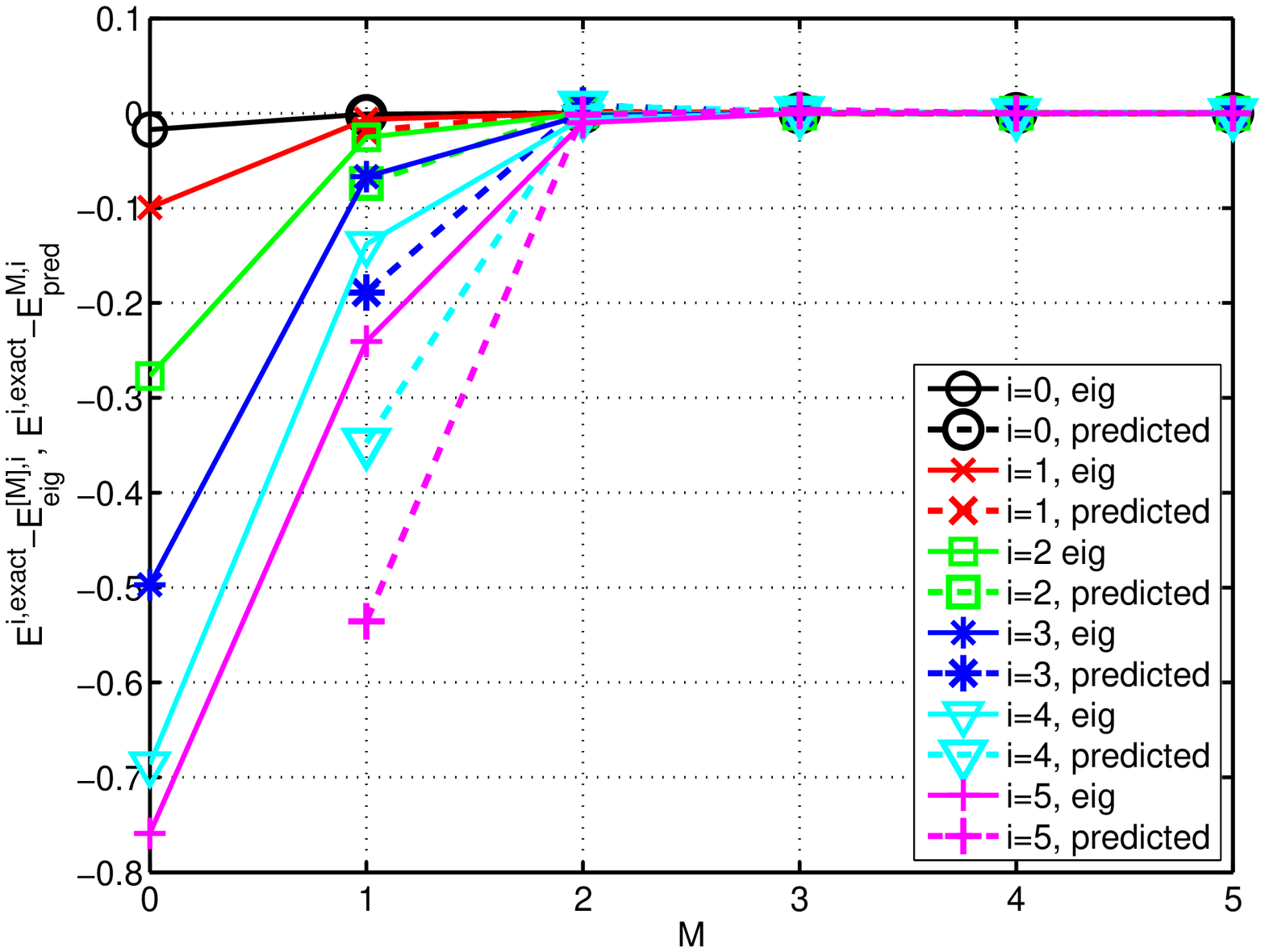}
  \includegraphics[height=6.5cm]{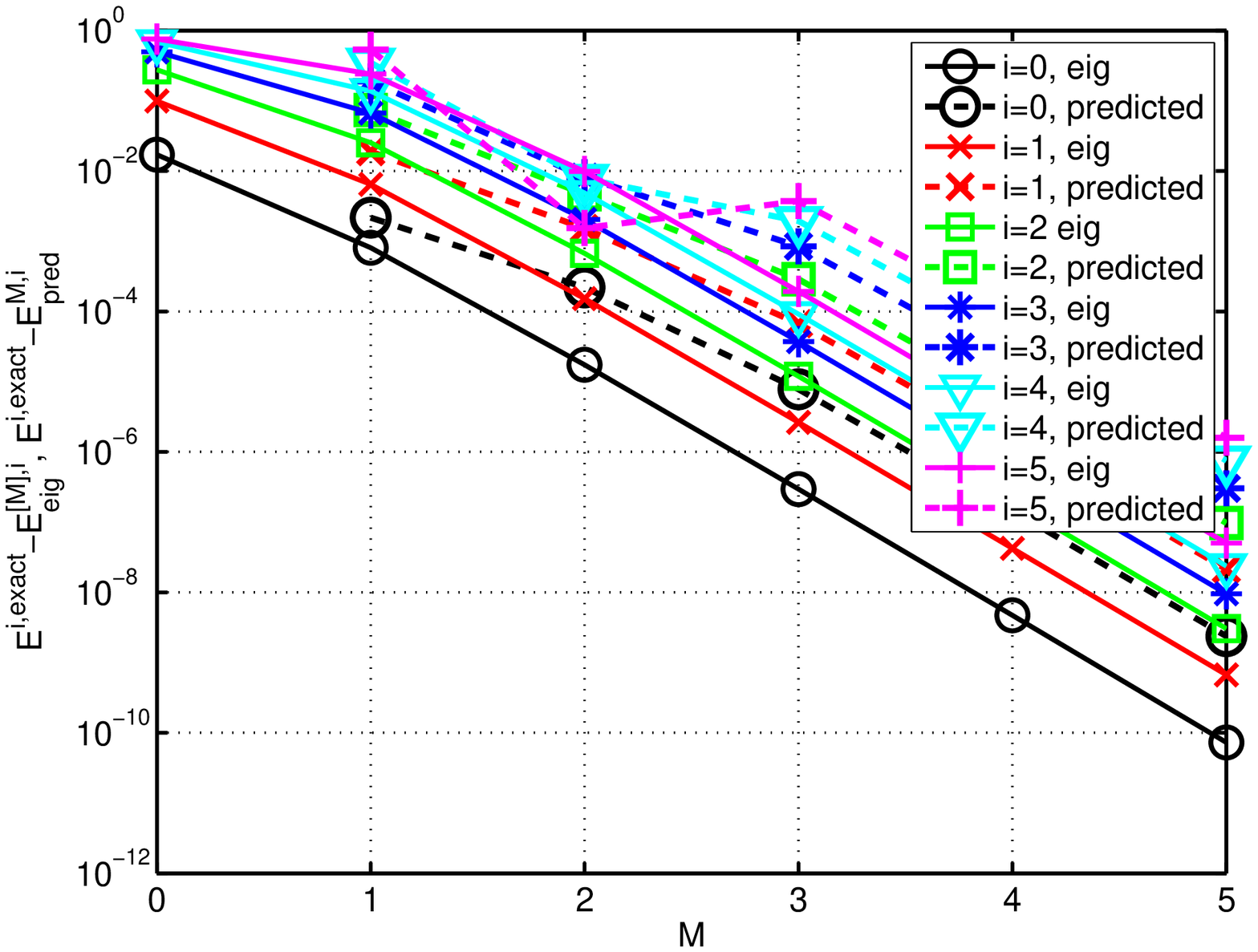}
  \caption{Energy differences vs. norm square differences of the $M$th level solutions of the eigenvalue equations and the predictions for the next resolution levels for the ground and excited states $i=0,\ldots,5$. A model system of a 1D harmonic oscillator and Daubechies basis set of support length 8 were applied. $E^{[M],i}$ denotes the energy levels calculated from the $M$th level eigenvalue equations and $E^{M+1,i}_{pred}$ means the $M+1$st level predictions calculated from the $M$th level solutions. Atomic units were used.}
  \label{fig:normenergdif}
\end{figure}

Due to the improved precision of the wave functions, the energy expectation values can be also be more precise. The exact energies of the harmonic oscillator's $i$th excited states can be compared to the eigenenergies $E^{[M]}$ of the Hamiltonian expanded up to the $M$th resolution level and to the expectation values $E^M_{pred}$ calculated with the $M$th level predicted wave functions, where the predictions are deduced from the $(M-1)$th level approximation. The energy level corrections resulting from the predicted wave functions are given in table~\ref{tab:energy}. The calculations were carried out on a simple, one-dimensional harmonic oscillator with Daubechies wavelets of support length 8. It can be seen, that the predicted energy mostly overcompensates the error of the energy arising from the solution of the previous resolution level eigenvalue equation.
\begin{sidewaystable}
\small
\centering 
  \caption{\ Energy levels of a simple 1D harmonic oscillator model system. The exact energies, the energies $E^{[M]}$ calculated as eigenvalues of the $M$th resolution level approximation of the Hamiltonian, and the predicted energies $E^M_{pred}$ calculated from $E^{[M-1]}$. Atomic units and Daubechies-8 basis set were used.}
  \label{tab:energy}
  \begin{tabular}{|l|c|c|c|c|c|c|}
  \hline
  Energy & Ground state & 1st excited state & 2nd excited state & 3rd excited state & 4th excited state & 5th excited state  \\ \hline
  Exact        & 0.5 & 1.5 & 2.5 & 3.5 & 4.5 & 5.5 \\
  $E^{[0]}$    & 0.517112256390810 &  1.599404458146794 &  2.777022029081063 &  3.997082442456408 &  5.186398997999037 &  6.259300101049636\\
  $E^1_{pred}$ & 0.502172810810787 &  1.518800774886581 &  2.573978630132646 &  3.689029670234657 &  4.847482745389376 &  6.035664236663306\\
  $E^{[1]}$    & 0.500808994455534 &  1.506441583382804 &  2.525266283013718 &  3.566883650097235 &  4.637885946573929 &  5.740594246712219\\
  $E^2_{pred}$ & 0.499779606006445 &  1.498478772113361 &  2.495343564778601 &  3.491846904710259 &  4.492240615775458 &  5.501541025240728\\
  $E^{[2]}$    & 0.500017441275289 &  1.500152737719495 &  2.500673509869070 &  3.502041349156252 &  4.504873856129472 &  5.509890004116425\\
  $E^3_{pred}$ & 0.499992232871423 &  1.499933033305139 &  2.499711780886731 &  3.499154925008878 &  4.498064719850611 &  5.496263440131640\\
  $E^{[3]}$    & 0.500000295257151 &  1.500002639582547 &  2.500011930589652 &  3.500037205990299 &  4.500091693592365 &  5.500192556080364\\
  $E^4_{pred}$ & 0.499999854828044 &  1.499998706339830 &  2.499994180696348 &  3.499981963890681 &  4.499955871984413 &  5.499908076893959\\
  $E^{[4]}$    & 0.500000004706870 &  1.500000042294175 &  2.500000192296497 &  3.500000603696133 &  4.500001498629072 &  5.500003171323336\\
  $E^5_{pred}$ & 0.499999997629393 &  1.499999978717204 &  2.499999903347566 &  3.499999697018824 &  4.499999249169179 &  5.499998414129641\\
  $E^{[6]}$    & 0.500000000072438 &  1.500000000664393 &  2.500000003027253 &  3.500000009522552 &  4.500000023679339 &  5.500000050205642\\
  $E^6_{pred}$ & 0.499999999962497 &  1.499999999662935 &  2.499999998466288 &  3.499999995179508 &  4.499999988018145 &  5.499999974608066\\
  $E^{[6]}$    & 0.500000000027025 &  1.500000000022808 &  2.500000000046947 &  3.500000000155680 &  4.500000000382068 &  5.500000000791360\\
  \hline
  \end{tabular}
\end{sidewaystable}
In Figure~\ref{fig:energdif} the errors of the energies $E^{[M]}$ and $E^{M+1}_{pred}$ are plotted for resolution levels $M=2,\ldots,6$, which shows a clear improvement of the predicted energies compared to the eigenvalues of the previous resolution levels.
If the energy differences versus the wave function norm differences are plotted on a logarithmic scale, we count on decreasing the energy errors approximately linearly as a function of the norm-square differences, i.e., a power law behavior is expected. Figure~\ref{fig:normenergdif} shows the energy difference as a function of the norm-square difference of the solutions of the eigenvalue equations at various resolution levels and the predictions arising from these solutions for the next resolution levels. The expected linear function with slope 1 almost fits the curves.

\section{Predictions for the 2nd finer resolution coefficients and higher resolution levels}

\begin{figure*}[hbt]
\centering
  \includegraphics[height=4.5cm]{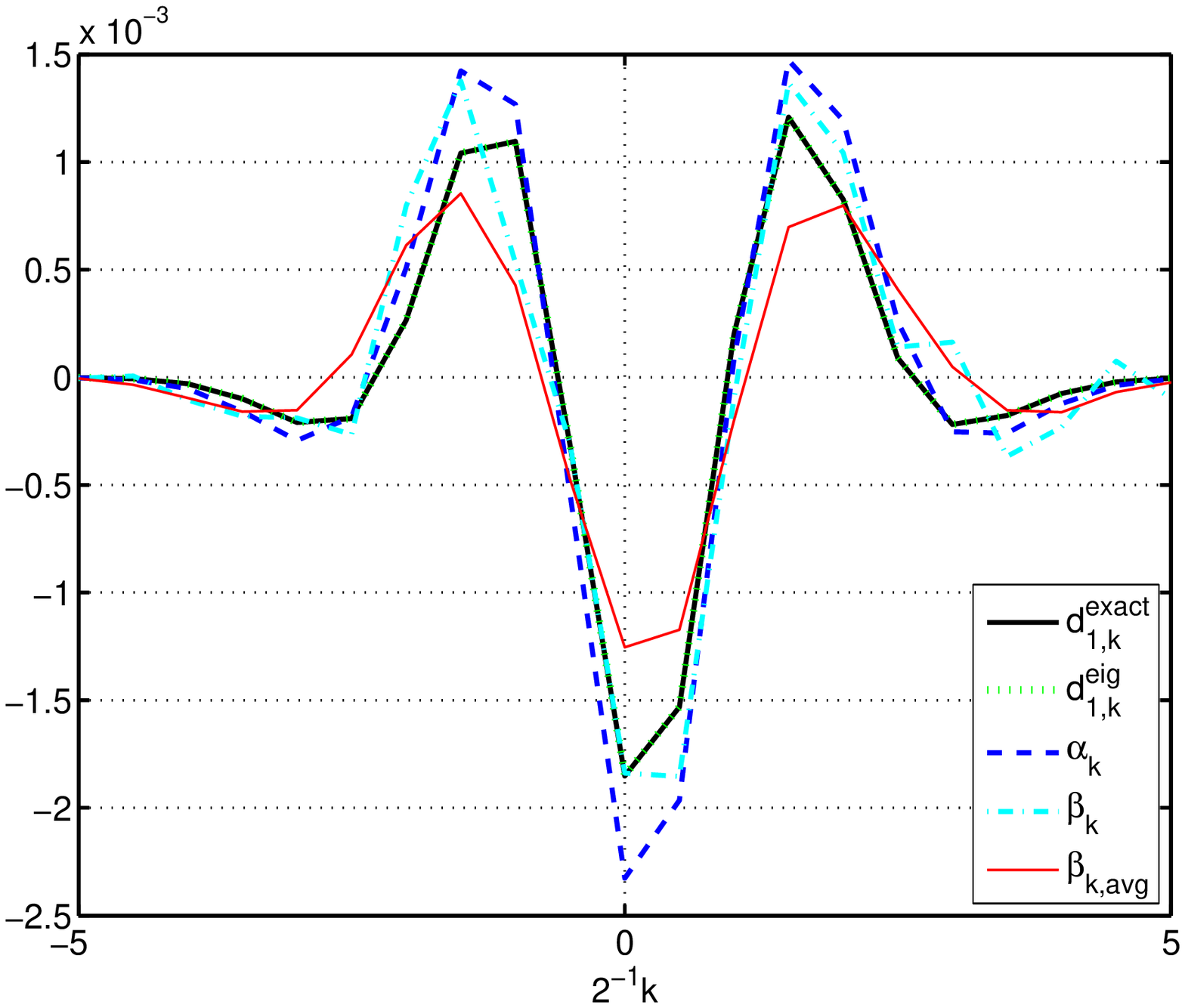}
  \includegraphics[height=4.5cm]{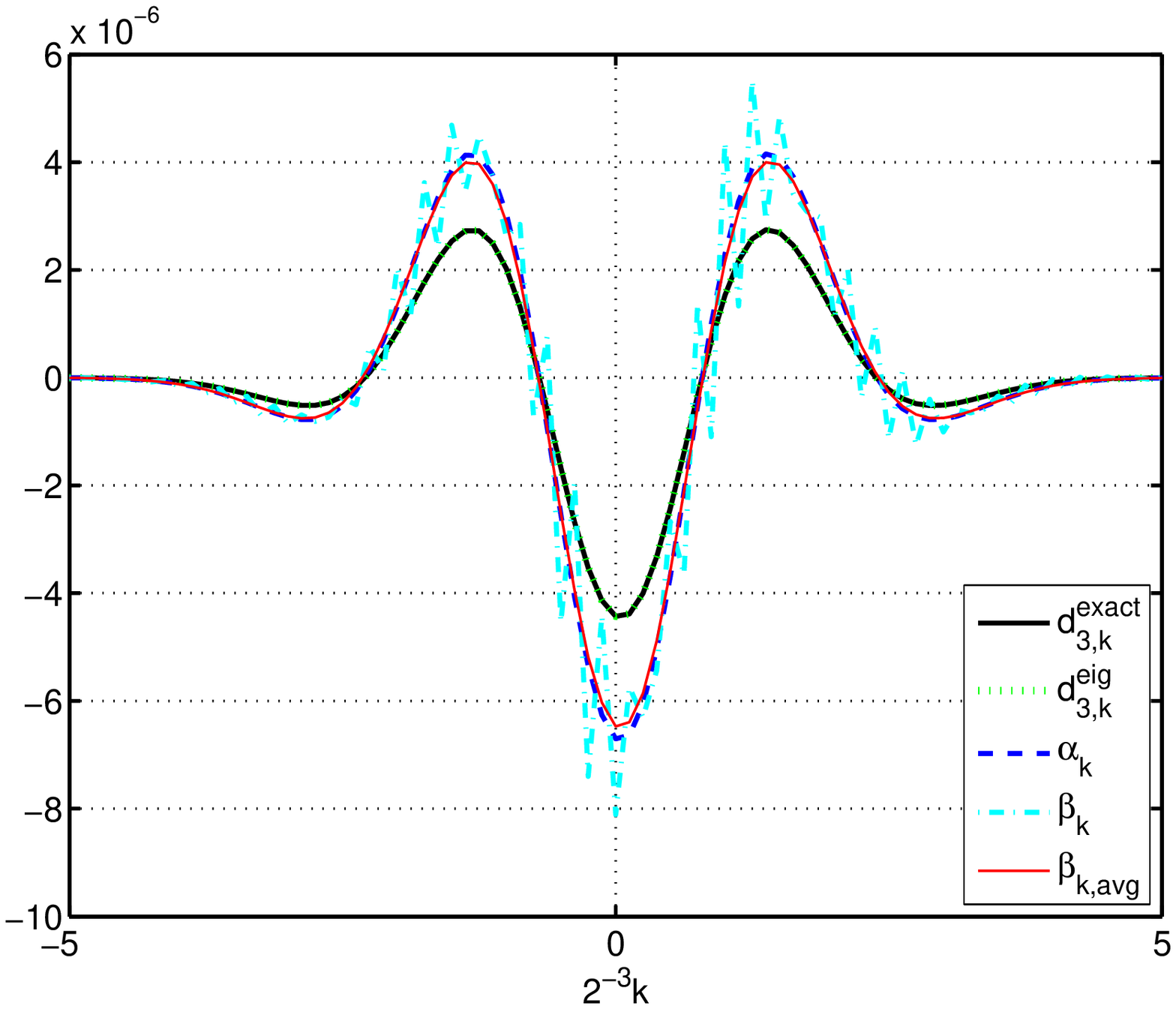}
  \includegraphics[height=4.5cm]{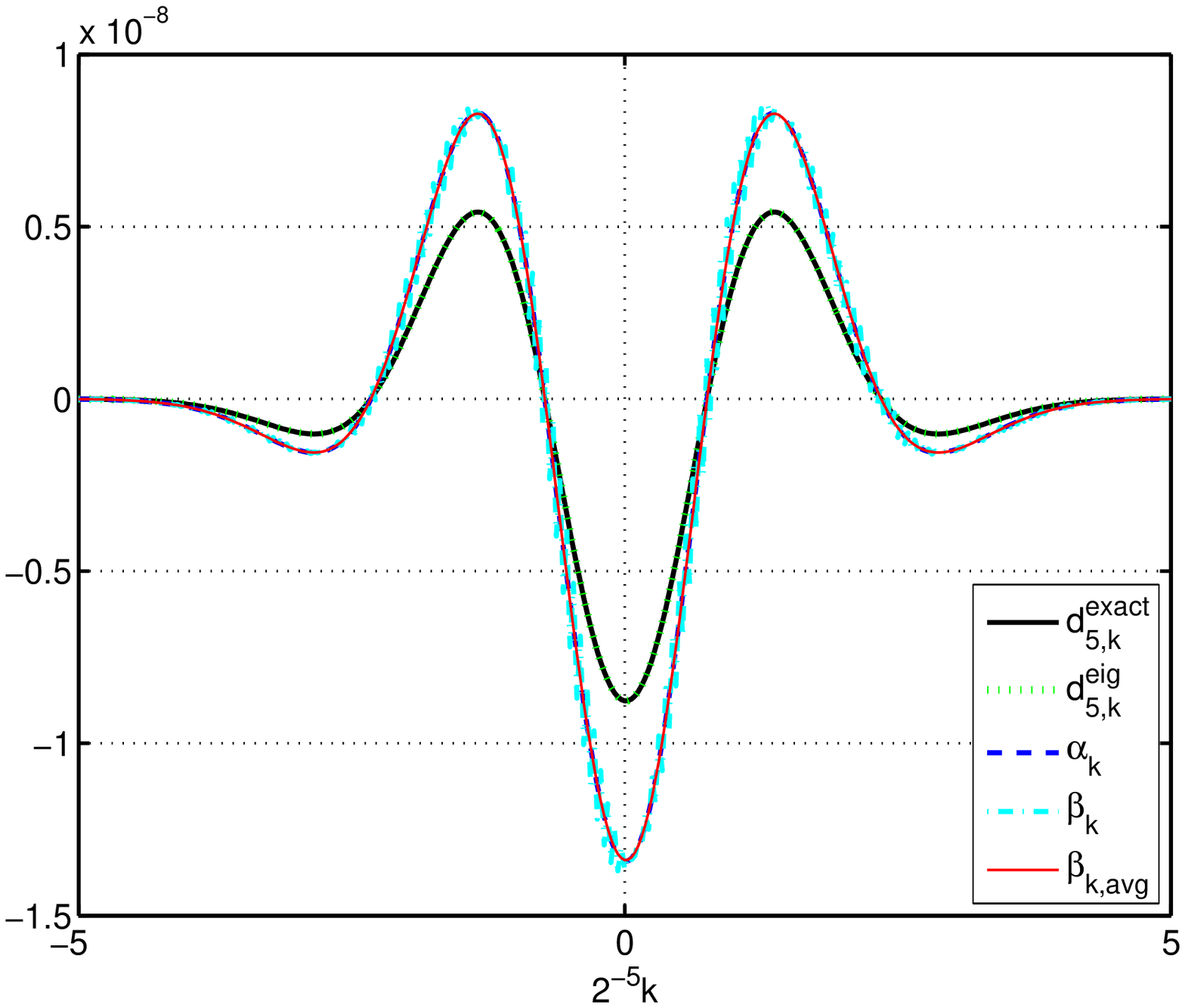}
  \caption{Ground state expansion coefficients $d^{exact}_{M,k}$, $d^{eig}_{M,k}$, predicted coefficients $\alpha_k$, and secondary predicted coefficients $\beta_k$ for a 1D harmonic oscillator model system using Daubechies-8 basis set for resolution levels $M=1,3,5$. Also an averaged secondary prediction is plotted, which gives a better approximation of the real coefficients. Atomic units were used.}
  \label{fig:ppred}
\end{figure*}

We have demonstrated\cite{jcc}, that a computationally economic calculation can predict the wave functions' next resolution level coefficients, but using these coefficients as a basis of another refinement could result in even more economic calculation scheme. Let us suppose, that we have an eigenvector of the $M$th resolution level problem $\Psi^{[M]}$, and the predicted wavelet coefficients $\alpha_k$ for the next resolution level, i.e., we have a predicted wave function
\begin{equation}\label{Psipred}
    \Psi^{M+1}_{pred}=\Psi^{[M]}+\sum_k \alpha_k w_{M,k}.
\end{equation}
Similarly to the first prediction, a secondary predicted wave function can be introduced by using one wavelet of the next resolution level,
\begin{equation}\label{Phi_1wavelet}
    \Phi^{M+2}_{pred}(\beta_k)=\Psi^{M+1}_{pred}+\beta_k\cdot w_{M+1,k}
\end{equation}
a new energy expression can be derived
\begin{equation}\label{Energy1}
    \mathcal{E}(\beta_k)=\frac{\langle \Phi^{M+2}_{pred}(\beta_k)|\hat{H}|\Phi^{M+2}_{pred}(\beta_k)\rangle}{\langle \Phi^{M+2}_{pred}(\beta_k)|\Phi^{M+2}_{pred}(\beta_k)\rangle}.
\end{equation}
Applying the variation principle, the resulting coefficients, similarly to (\ref{alfa}) are
\begin{equation}\label{beta}
  \beta_k=\left\{\begin{array}{cc}
    -\mu+\sqrt{\mu^2+1}, &\quad\mbox{if } \langle w_{M+1, k}|\hat{H}|\Psi^{M+1}_{pred}\rangle> 0\\
    -\mu-\sqrt{\mu^2+1}, &\quad\mbox{if } \langle w_{M+1, k}|\hat{H}|\Psi^{M+1}_{pred}\rangle< 0\\
    0                    &\quad\mbox{if } \langle w_{M+1, k}|\hat{H}|\Psi^{M+1}_{pred}\rangle=0
  \end{array}\right..
\end{equation}
Here, the shorthand notation covers a bit more complicated meaning
\begin{equation}\label{lambda}
  \mu=\frac{E^{M+1}_{pred}-\langle w_{M+1, k}|\hat{H}|w_{M+1, k}\rangle}{2\langle w_{M+1, k}|\hat{H}|\Psi^{M+1}_{pred}\rangle}.
\end{equation}
As it can be seen from the previous two sections, the energy level $E^{M+1}_{pred}$ differs not too much from the previous energy $E^{[M]}$, moreover, after a given, but not too large resolution level, the complete term $E^{M+1}_{pred}$ is negligible compared to the other component of the denominator in $\mu$. Using the same model system as previously, in Figure~\ref{fig:ppred} we have plotted the exact $M$th resolution level expansion coefficients $d^{exact}_{M,k}$, the eigenvectors of the $M$th level Hamiltonian $d^{eig}_{M,k}$, the first predicted coefficients $\alpha_k$, arising from the $(M-1)$th level solution, and the secondary predicted coefficient $\beta_k$, which is derived from the $(M-2)$th level eigenvector solution, and from its prediction to level $M+1$, i.e., from $\Psi^{M-1}_{pred}$. Note, that the coefficients of level $m<M$ also differ for the various ``solutions'' of level $M$, however, they are not plotted due to only slightly visible differences.

The figure shows, that the secondary predicted coefficients have a slight oscillation around a rather good prediction for the exact values of the coefficients. This oscillatory behaviour can be reduced by an averaging of the neighbouring coefficients (see the thin line on the figures), however, this aspect needs further investigation.

\section{Summary}

We have presented a prediction method for the magnitude of the next resolution level coefficients of a wavelet based electron structure calculation, if an eigenvalue calculation of the discretised Hamiltonian at resolution level $M$ is already carried out. We have given the scaling behaviour of the predicted coefficients, which was $2^{-6M}$ and the total omitted weight if the calculation is stopped at resolution level $M$.

We have demonstrated, that the energy expectation values with the predicted wave functions give better result than the original energies, however, an overcompensation of the errors occurs often.

We have also studied the secondary predicted coefficients, that give in average a fair approximation of the real wavelet expansion coefficients, even though an oscillation around the ideal value and a slight overestimation can also be experienced in our model system.

\end{document}